# Transparent and flexible field-effect transistors and mem-transistors with electroactive layers of solution-processed organic polyradicals

Deepa Singh[1,2], François Magnan[2,3], Joe B. Gilroy[2,3] *, and Giovanni Fanchini[1,2,3] *

[1]Department of Physics and Astronomy, The University of Western Ontario, 1151 Richmond St., London, Ontario N6A 3K7, Canada.

[2]Centre for Advanced Materials and Biomaterials Research (CAMBR), The University of Western Ontario, 1151 Richmond St., London, Ontario, N6A 5B7, Canada.

[3]Department of Chemistry, The University of Western Ontario, 1151 Richmond St., London, Ontario, N6A 5B7, Canada.

*Solution-processed polymers are at the core of organic electronics.*[1-5] *Polyradicals –polymers in which each repeating unit contains an unpaired spin– are unique alternatives to their π-conjugated, semiconducting counterparts.*[6,7] *Unique of polyradicals are tunable charge states localized at their repeating units, which enable electrically switchable charge transport regimes.*[8] *Tremendous efforts were focused on polyradical memristors*[8-9] *and batteries.*[10] *Notwithstanding recent progress in doping,*[11] *polyradical field-effect transistors (PR-FETs) have not been reported. Here, we show that vertical architectures, with drain-source contacts sandwiching the active layer of a strongly correlated 6-oxoverdazyl polyradical,*[12] *leads to on/off ratios >$10^3$ in p-type PR-FETs. Hole injection occurs via contact doping by tunable charge states at the polyradical-electrode interface. Transparent and flexible PR-FETs are also reported. PR-FETs are superior to existing organic FETs as they combine memristor and transistor functions in one mem-transistor device, offering a unique potential for the circuital simplification of organic electronics.*

Numerous studies on electronic-grade organic polymers have focused on enhancing their performance by studying the effect of solvents,[13-14] morphology,[15] metal-polymer interfaces,[16-17] and doping.[18-19] However, the vastest majority of organic FETs reported rely on π-conjugated semiconducting polymers as the electroactive layer. Even though polyradicals successfully compete with these polymers in several critical applications,[8-10] their role in transistors has been insofar limited to the functionalization of multi-component devices. The inclusion of poly(2,2,6,6-tetramethyl-piperidinyl-oxy-4-yl-methacrylate) in pentacene FETs enhanced their on/off ratio.[20] Similar effects were obtained by functionalizing carbon nanotube FETs with poly(2,6-di*tert*-butyl-4-(3,5-di-*tert*-butyl-4-phenoxyl)(4-vinylphenyl-methylene-cyclohexa-2,5-dienone),[21] in a process relatively similar to nanotube functionalization by diazonium salts.[22] Although these reports illustrate the benefits of polyradical incorporation in electroactive layers of other materials, FETs with polyradical thin films as electroactive layers have not yet been constructed.

We present a simple route towards polyradical layers of transistor-grade quality from the processing of poly-[1,5-diisopropyl-3-(*cis*-5-norbornene-*exo*-2,3-dicarboxiimide)6-oxoverdazyl] (P6OV), a nonconjugated 6-oxoverdazyl polyradical.[8,12] Fig. 1a-c demonstrate P6OV in the three charge states in which its radical monomers exist. The density-of-states of P6OV near the Fermi level contains bonding $\pi$ electrons in the highest occupied molecular orbitals (HOMO, $\pi$), unpaired electrons found in singly-occupied molecular orbitals (SOMO), and empty lowest unoccupied molecular orbitals (LUMO, $\pi^*$). The HOMO-LUMO ($\pi$-$\pi^*$) bandgap is 4.6 eV.[8] As P6OV is nonconjugated, unpaired electrons are localized at each 6-oxoverdazyl ring. This causes them to sit at varying energies within the bandgap, depending on their charge state (Fig. 1a-c). Three states are possible: neutral, cationic, and anionic. The neutral SOMO lies at mid-gap. Cationic states sit at lower energy (1.5 eV above the HOMO, Fig. 1b) than their SOMO counterparts, while anionic states sit higher (3.1 eV above the HOMO, Fig. 1c). This is due to strong Coulombic repulsion and correlation energy in few-electron systems, which makes doubly-occupied, anionic states less stable than their cationic or neutral counterparts. These charge-tunable states lead to two switchable transport regimes: semiconducting ("0") with P6OV mostly in neutral state, and semimetallic ("1") with an alternation of cations and anions, stabilized by electrostatic interaction.[8]

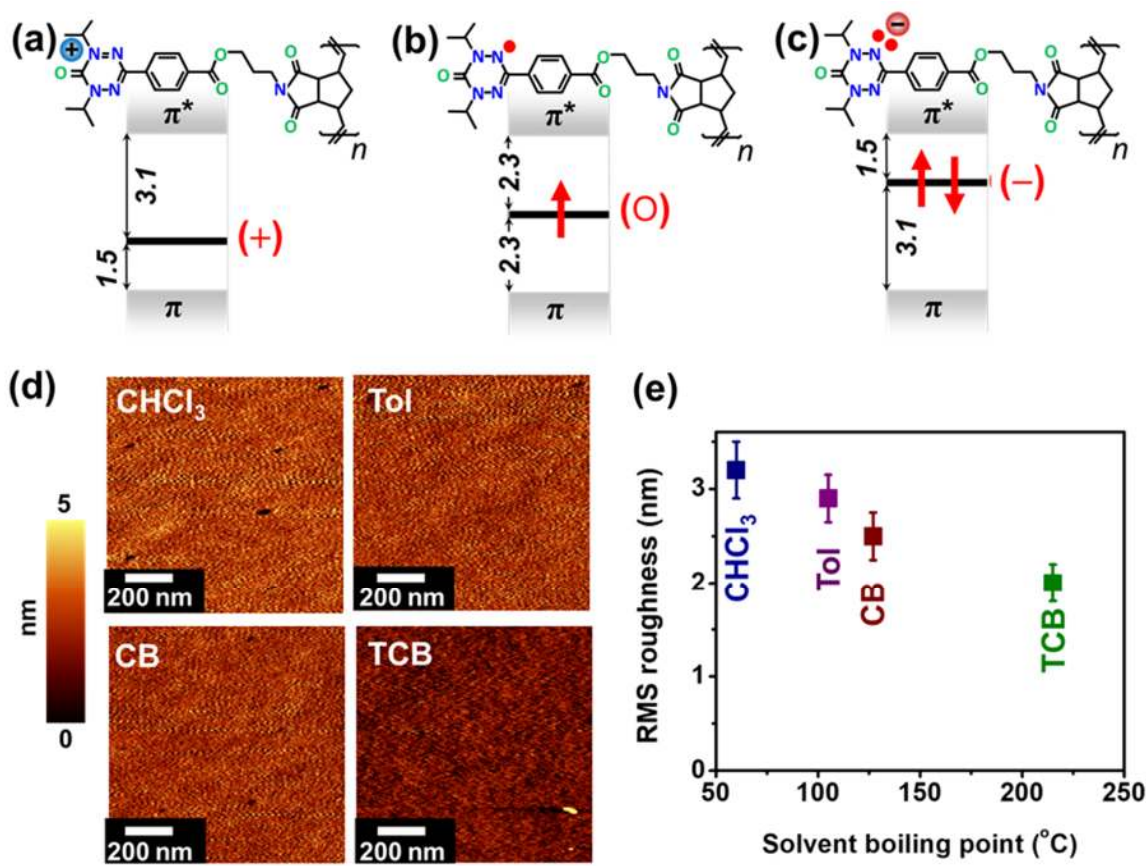

**Fig. 1 | Electronic structure and morphology of P6OV thin films – a.** *Chemical and electronic structure of P6OV in cationic,* **b.** *neutral, and* **c.** *anionic charge state.* **d.** *AFM of P6OV thin films spun from different solvents. From lower to higher boiling point (BP): chloroform (CHCl$_3$, BP = 61°C), toluene (Tol, 110°C), chlorobenzene (CB, 132°C) and 1,3,5-trichlorobenzene (TCB, 208°C).* **e.** *Root mean square (RMS) film roughness decreasing at increasing BP. All films are of ~50 nm thickness.*

Because 6-oxoverdazyl polyradicals are normally synthesized in the neutral form,[12] P6OV offers high stability when dissolved in a variety of nonpolar solvents. Solvent choice is vital to achieve optimal film-forming properties in many deposition techniques for organic electronics,[23] for example spin coating. We have spun P6OV thin films from a variety of solvents at different boiling points. Fig. 1d shows typical atomic force micrographs (AFM) of ~50-nm thick P6OV films, demonstrating remarkable smoothness even at relatively low thicknesses, and regardless of the processing solvent. Surface roughness (Fig. 1e) was < 3.5 nm in all cases, with the highest boiling solvents yielding to the smoothest films, presumably because of slower solvent evaporation and nucleation rate. Extreme smoothness is often necessary to achieve void-free active layers, low concentrations of charge-trapping sites, and high mobility suitable for transistor operation at low enough gate voltages.[15]

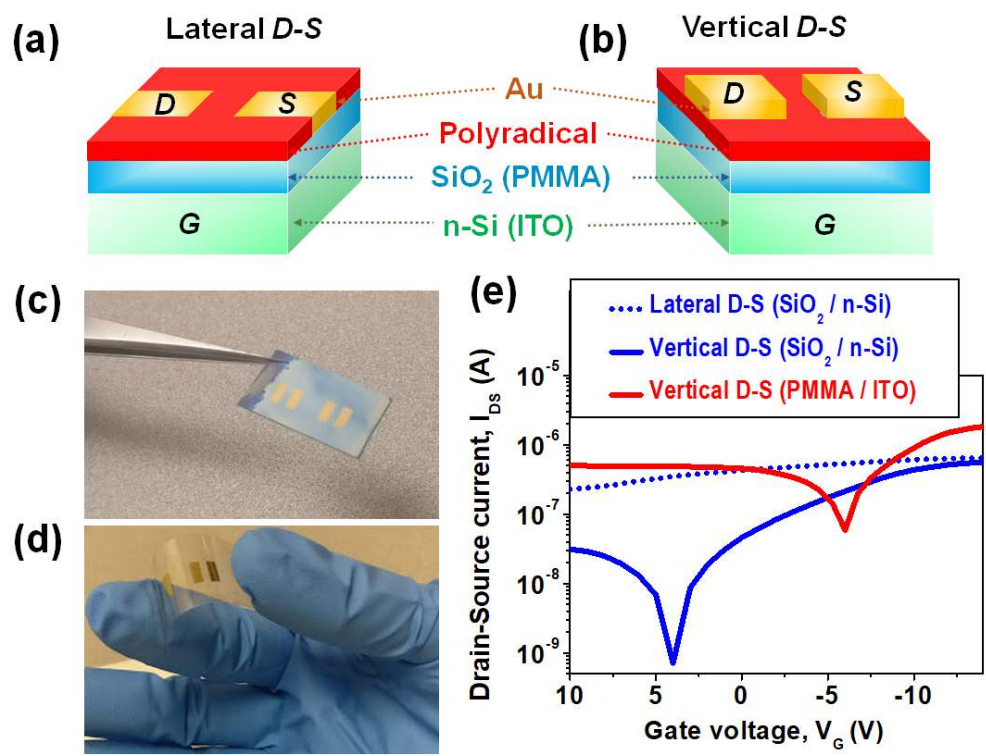

**Fig. 2 │ Device architecture and its influence on PR-FET transfer characteristics - a.** *Lateral-contact, and* **b.** *Vertical-contact architectures, with the former more commonly used in organic electronics, but leading to modest gate effects in PR-FETs. Devices fabricated on* **c.** *SiO₂/n-type Si, and* **d.** *flexible polyethylene teraphtalate, with transparent indium tin oxide (ITO) gate and polymethyl-methacrylate (PMMA) gate insulator.* **e.** *Gate transfer curves ($I_{DS}$–$V_G$) for different substrates and architectures. Higher $I_{ON}/I_{OFF}$ ratios are evident in vertical architecture, along with ~500-time higher leakage currents in the flexible device with PMMA gate insulator.*

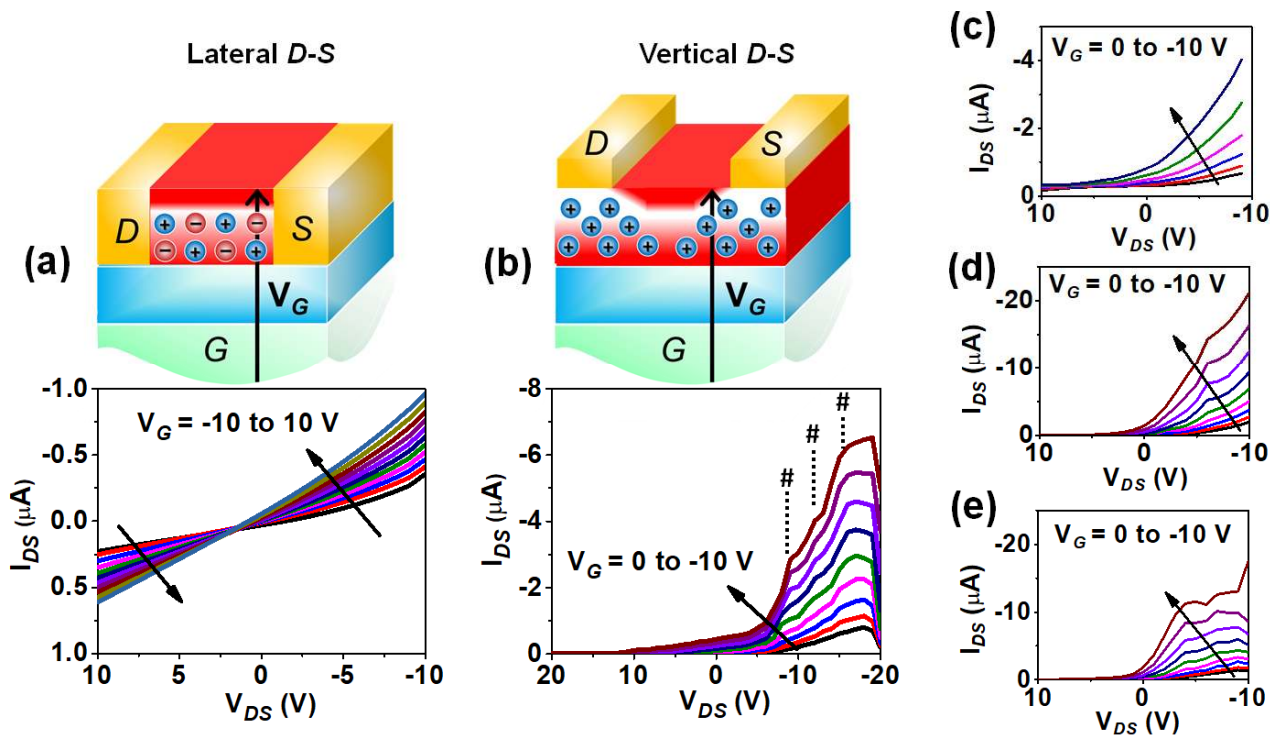

**Fig. 3 | Influence of morphology on PR-FET output characteristics - a.** $I_{DS}-V_{DS}$ *output in the lateral-contact architecture with ambipolar characteristics, consistent with semimetallic transport properties of P6OV in the "1" regime and spun from TCB.* **b.** $I_{DS}-V_{DS}$ *output in the vertical-contact architecture with p-type characteristics, consistent with semiconducting transport properties of P6OV in the "0" regime and spun from TCB. "Kinks" corresponding to $I_{DS}$ current surges are indicated with (#). Vertical-contact transistors spun from* **c.** *$CHCl_3$,* **d.** *Tol, and* **e.** *CB, showing the gradual quenching of "kinks" when low-boiling solvents (e.g. $CHCl_3$) are utilized for spin coating.*

We compared PR-FETs fabricated in two distinct architectures. In the lateral-contact architecture (Fig. 2a) commonly used in organic electronics,[24] drain and source (D-S) gold contacts were pre-grown on the gate insulator prior to the deposition of the polyradical layer. In the vertical-contact architecture (Fig. 2b), D-S contacts were grown onto the polyradical film that was sandwiched between the gate insulator and the contacts, with extreme attention to avoid sample heating during contact growth. Fig. 2c-d depict PR-FETs on opaque/rigid and transparent/flexible substrates. Fig. 2e compares the gate transfer curves from different substrates, as well as lateral and vertical-contact architectures. Significant on/off ratios and strong *p*-type character are noticeable, as opposed of modest dependency of D-S current on the gate voltage ($V_G$) in the lateral-contact architecture. Although flexible PR-FETs on plastics exhibit larger leakage currents than their rigid counterparts, they showed reproducible operation after repeated folding, with potential applications in driving flexible liquid crystal displays and wearable electronics.[2]

Drain-source current ($I_{DS}$) vs. voltage ($V_{DS}$) output characteristics were measured for both architectures. Fig. 3a demonstrates the $I_{DS}$–$V_{DS}$ output in the lateral-contact architecture. These PR-FETs exhibit ambipolar behavior, with comparable gating effects at both positive (electron-related) and negative (hole-related) $I_{DS}$. On/off ratios ($I_{ON}/I_{OFF} \approx 3$) are modest in this architecture. Such findings are consistent with the semimetallic transport properties found by Ezugwu *et al.*[8] for P6OV memristors in the high-conducting ("1") regime. In this regime, repeating units are alternating anions or cations. As the Fermi level sits at mid-gap, this produces injection of similar rates of opposite charges at both electrodes, with similar densities of positively and negatively charged monomers at $V_G = 0$. Ambipolar behavior is also consistent with the ambipolar redox properties previously observed by Paquette *et al.*[12] via cyclic voltammetry.

To achieve meaningful on/off ratios, PR-FETs in the vertical-contact architecture were constructed (Fig. 3b). Vertical PR-FETs exhibit $I_{ON}/I_{OFF} > 10^3$ and *p*-type characteristics. Such properties are consistent with the semiconducting properties found by Ezugwu *et al.*[8] in P6OV memristors in the low-conducting ("0") regime. In this regime, most of the radical monomers are in the neutral state, and charge transport occurs via diffusion of carriers localized at specific repeating units.[8] These data demonstrate that, in PR-FETs with gold D-S contacts, majority carriers are holes. As no differences were introduced other than contact design, origin of the hole current must be assigned to the role played by the P6OV layer sandwiched between the D-S contacts and the gate insulator, in a process of doping-by-contact. A similar process in a vertical π-conjugated FET architecture is described by Günther *et al.*[18] It is noteworthy that P6OV enables a remarkable simplification over such architectures, with one less fabrication step and, differently from ref. 18, no additional semimetallic interlayer required for hole injection and contact doping. This is an anticipated consequence of P6OV functioning both as the semiconducting active layer (in the "0" regime) and the semimetallic contact doping layer (in the "1" regime).

Additional evidence of contact doping can be inferred from the role played in the transistors' electrical characteristics, by the surface roughness of the active layer. Fig. 3b-e show that the $I_{DS} - V_{DS}$ output characteristics of PR-FETs are often not smooth: specific discrete values of $V_{DS}$ (indicated with # in Fig. 3b), produce drain current surges, corresponding to "kinks" in the $I_{DS} - V_{DS}$ outputs. In the literature, similar "kinks" are linked to deep-level charge-trapping sites at the semiconductor interfaces.[26-27] If corroborated by future experiments, such assignment will further support the critical role of interfaces on the hole injection processes in PR-FET channels. A good correlation exists between the intensity of "kinks" (Fig. 3b-e) and the surface roughness measured by AFM (Fig. 1d) in P6OV layers spun from solvents at varying boiling point. Fig. 3b-c demonstrates that drain current surges are more prominent in the smoothest PR-FET active layers that are solution-processed from high-boiling solvents. "Kink" intensity increases upon annealing at 120°C in inert atmosphere (see Supplementary Information) in a procedure designed to eliminate trapped solvent from the active layer. No "kinks" are noticeable in PR-FETs spun from $CHCl_3$, the lowest-boiling solvent among those reported in Fig. 1e, which corresponds to the least smooth active layer (and the highest surface density of contacts). These findings point to the major role played by interface roughness in polyradical contact doping.

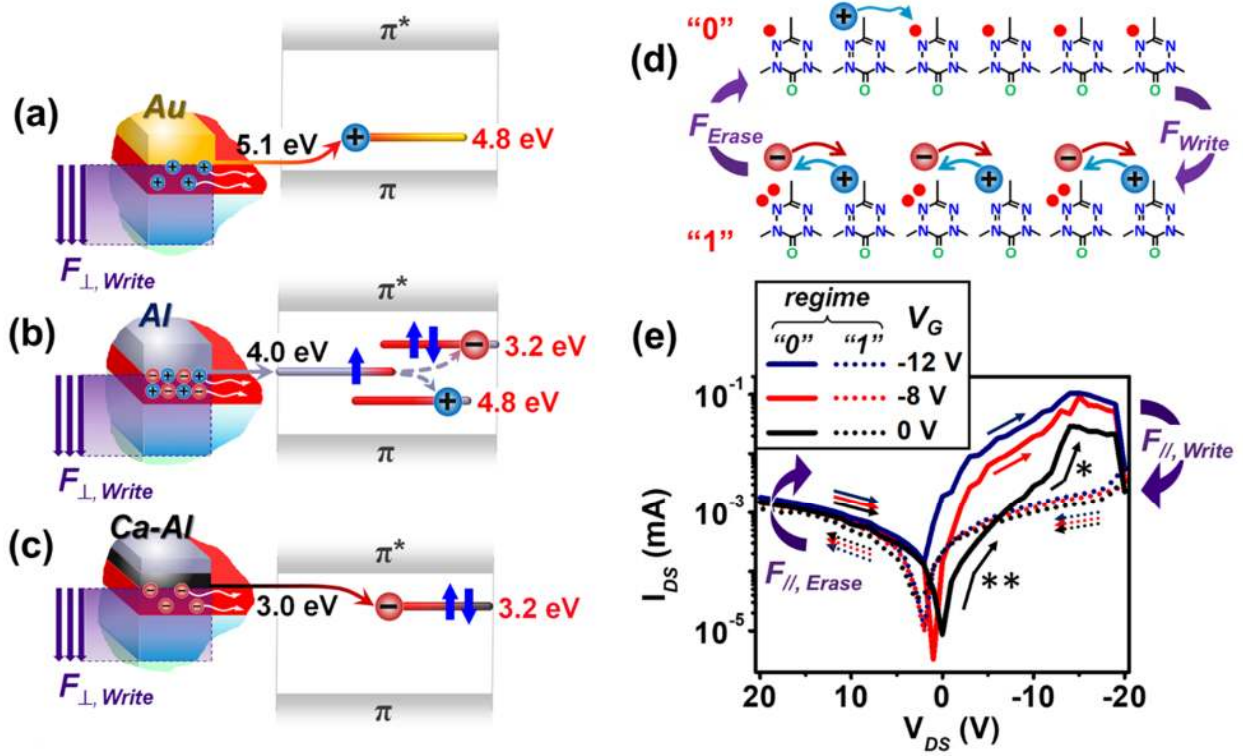

**Fig. 4 │ Contact doping and PR-FET operation as mem-transistors - a.** *Contact doping process with Au contact at ~5.1 eV work function, enabling selective hole injection from the source to the P6OV layer under the contact when a sufficient gate voltage, leading to electric field $F_\perp > F_{\perp, Write}$ is applied;* **b.** *Same as above with Al contact at ~4.0 eV work function, enabling both electron and hole injection; and* **c.** *Proposed contact doping with Al/Ca electrodes at ~3.0 eV work function, enabling selective electron injection. In all cases a-c carriers are generated in the injection layer below the contacts and diffuse in the channel.* **d.** *Schematic of "0" and "1" electrical transport regimes with simplified 6-oxoverdazyl structures,[8] and* **e.** *Mem-transistor with Au contacts cycles at different gate voltages, with device switching from gate-modulated "0" regime to gate-independent "1" regime. Direction of cycles is indicated by the arrows. The "0" regime may lead to either (\*\*) $I_h^{(0)} < I_h^{(1)}$ or (\*) $I_h^{(0)} > I_h^{(1)}$ depending on $N_h^{(0)}$ that also depends, at high enough voltages, on $V_{DS}$ due to superlinear Poole-Frenkel effect.*

To elucidate the origin of *p*-type Au-contact doping, PR-FETs with metallic contacts other than Au were considered (see Supplementary Information). Fig. 4a-c reports their work functions as well as their alignment with the cationic, neutral, and anionic states in P6OV.[8] The work function of Au ($W_f \approx 5.1 \pm 0.2$ eV) sits well below the 4.8 eV cationic level of P6OV. Thus, Au preferentially injects holes over electrons into 6-oxoverdazyl radical monomers. Negative gate voltages generate off-plane electric fields ($F_\perp$) in P6OV under the contacts. If $F_\perp \leq F_{\perp, \text{Write}}$, P6OV switches into the "1" transport regime in the injection layer sandwiched between the contact and the gate insulator. This is consistent with what happens in two-terminal P6OV flash memristors,[8] where equal amounts of alternating anions and cations are thus produced (Fig. 4d). However, because of the preferential injection of holes from gold in PR-FETs, increasing cation monomer concentrations are generated at increasing negative $V_G$. This results in gate-modulated hole diffusion through the channel, and in the *p*-type nature of the devices.

In the case of aluminum ($W_f \approx 4.0 \pm 0.3$ eV), $W_f$ aligns to the neutral SOMO state of P6OV (Fig. 4b). Thus, aluminum transfers equal rates of electrons and holes to the polyradical injection layer, which explains why Al-contacted PR-FETs are ambipolar even in the vertical-contact architecture (see Supplementary Information). A continuation to our work will be the demonstration of *n*-type PR-FETs. Hybrid Al/Ca contacts ($W_f \approx$ 2.8-3.2 eV) preferentially inject electrons into the 3.2 eV anionic level of P6OV (Fig. 4c).[8] However, given the electron-rich nature of P6OV polyanions, these are highly susceptible to oxidation and decomposition. This explains why we have not yet obtained *n*-type contact doping of P6OV at this time. Design of *n*-type PR-FETs may require polyanions with energy levels deeper than P6OV, and less prone to oxidation.

To further clarify electrical transport in Au-contacted vertical PR-FETs, hole/electron (h/e) mobilities $\mu_{h/e}^{(j)}$ were calculated in both regimes ($j$ = "0" or "1") from the device transconductances:

$[\mu_h^{(1)}\ \mu_h^{(0)}] = [2.3 \times 10^{-4}\ 4.4 \times 10^{-1}]$ cm$^2$/V/s for holes, and $[\mu_e^{(1)}\ \mu_e^{(0)}] = [2.9 \times 10^{-4}\ 0.8 \times 10^{-1}]$ cm$^2$/V/s for electrons. In the linear FET approximation,[28] $I_{DS}$ is primarily determined by the majority carrier (hole) current, $I_h^{(j)}$, according to Ohm's law:

$$I_{DS} \approx I_h^{(j)} \approx e \cdot N_h^{(j)} \cdot \mu_h^{(j)} \cdot V_{DS}, \qquad (1)$$

where $N_{h/e}^{(j)}$ indicates the hole/electron concentration and $e$ the elementary charge. Eqn. (1) shows that: *i*) in the "1" regime, low $\mu_h^{(1)}$ implies higher carrier concentrations, consistent with frequent carrier collisions, $N_h^{(1)} \approx N_e^{(1)}$, and with both carrier types relatively insensitive to gating; while *ii*) in the "0" regime, high $\mu_h^{(0)}$ implies lower carrier concentrations, consistent with more infrequent collisions, and with $N_h^{(0)}$ tunable via gate modulation. Specifically,[28] $N_h^{(0)}$ increases exponentially at higher negative $V_G$ and, because hole mobility is ~500 times higher in the "0" regime, $I_h^{(0)}$ may become *higher* than $I_h^{(1)}$, even though $N_h^{(0)}(V_G) \ll N_h^{(1)}$. This can be seen in Fig. 4e at both $V_G = -8$ V and $-12$ V. Additionally, the $I_{DS}$–$V_{DS}$ outputs are superlinear due to Poole-Frenkel effect.[8] At $V_G = 0$, this produces a crossover between "0" and "1" outputs, for which $I_h^{(0)} < I_h^{(1)}$ at the low negative $V_{DS}$ indicated with (**) in Fig. 4e, but $I_h^{(0)} > I_h^{(1)}$ at high negative $V_{DS}$, indicated with (*).

Collectively, Fig. 4e shows that such a crossover has profound implications in Au-contacted vertical PR-FETs. Application of high $|V_{DS}|$ lead to in-plane electric fields ($F_{//}$) that may be high enough to produce "0" to "1" switching and *vice versa*, not only under the contacts, but also inside the channel. Write ($\leq -20$V) and erase ($\geq 20$V) voltages are higher in-plane than off-plane. However, the $I_{DS}$–$V_{DS}$ characteristics still exhibit memory effects reminiscent of P6OV flash memristors.[8] Thus, PR-FETs effectively operate as mem-transistors above the $|V_{DS}| \geq 20$ V range,[29] while normal transistor operation occurs below such range. Basically, PR-FETs may integrate both transistor and memristor functions within one component and become either gate-modulated or gate-insensitive depending if they have been pre-programmed via "write" or "erase" D-S voltages.

In conclusion, we have shown that "doping-by-contact" of a 6-oxoverdazyl polyradical may lead to *p*-type PR-FETs with $I_{ON}/I_{OFF}>10^3$. Similar transistors were also constructed on transparent and flexible plastic substrates. A competitive advantage of PR-FETs over state-of-the-art organic FETs is that they can effectively function as mem-transistors when operated over the $|V_{DS}| \geq 20$ V range. Organic mem-transistors have been long-awaited technology since the first organic FET was developed by Tsumura *et al.*[30] in 1986. Inorganic mem-transistors based on 2D materials were heralded as an essential milestone towards the circuital simplification of inorganic electronics, specifically for data storage.[29] To date, no equivalent solution exists in organic electronics, arguably because state-of-the-art semiconducting polymers do not offer switchable electronic states with significant mobility. Our PR-FETs will represent an attractive solution towards the simplification of circuital organic electronics.

**Methods**

PR-FETs were assembled on (100) n-Si/SiO$_2$ (100 nm) wafers (MTI Corp. Ltd.) and, in their transparent and flexible version, on polyethylene terephthalate (PET) sheets coated with ITO at 100 Ω/sq resistivity (Sigma Aldrich). PET/ITO was coated by PMMA (Goodfellow Inc.) as a gate insulator. PMMA solution in CB (15 mg/mL) was stirred for 6 h at 50°C and spun at 1000 RPM to obtain a 100-nm thick insulating layer. P6OV was prepared by ring-opening metathesis polymerization (ROMP) as described elsewhere.[12] As previously described, extensive tests were carried out to confirm identity and purity, and to ensure the presence of one spin per monomer.[12] P6OV in powder form was dissolved in the requisite solvent, and subsequently deposited at varying speeds by spin coating from different solvents (TCB, CB, Tol and CHCl$_3$).

All P6OV thin film deposition, PR-FET device fabrication and electrical characterization were carried out in the same nitrogen-loaded glovebox (Nexus II, Vacuum Atmospheres Co.) with oxygen content <15 ppm and moisture level <3 ppm, equipped with a spin coater (KW4, Chemat Technologies Inc.) thermal evaporator (KJ Lesker Inc.) and testing probe station. Anhydrous solvents were obtained from Sigma Aldrich at the maximum available purity. Solutions of P6OV at different concentrations (5 and 10 mg/mL) were prepared. These solutions were stirred overnight in the glove box and filtered through 0.45-μm pore-diameter syringe filters prior to spin coating. Typically, at 10 mg/mL, 50-nm thicknesses were obtained from speeds of about 1000 RPM. The thermal evaporator (KJ Lesker Inc) was directly accessed from the glove box for the deposition of contacts through patterned shadow masks at $l_{DS}$ = 25 μm channel length and $w_{DS}$ = 2 mm channel width. Metal pellets for thermal evaporation were obtained from KJ Lesker Inc (Al and Ca) and Sigma-Aldrich (Au) under inert atmosphere and were also loaded into the thermal evaporator directly from the glove box. Special attention was paid to avoid device overheating during thermal evaporation, with metal evaporation temperatures lower than 75°C, as measured at the substrate using a calibrated K-thermocouple. Thicknesses of the S-D contacts were measured *in situ* using a computer-controlled Sycom STM2 quartz crystal thickness monitor. Thicknesses of P6OV thin films were obtained *ex situ* by AFM (Witec Alpha 300S microsvcope) on samples identical as those used for device fabrication. The same AFM instrument was used in tapping mode to image the surface of the samples.

Transfer and output characteristics of the devices were measured in a built-in probe station inside the glove box using two IEEE-488 computer-interfaced Keithley 2400 source meters, automated with a custom-built Matlab™ routine. Sample terminals were contacted using Signatone S725 probes. One source meter was used to sweep the gate voltage, while the second was connected to the D-S contacts. Multiple sweeps with polarity switching of the source meters were used to test the PR-FETs in mem-transistor operation mode. Carrier mobility was derived from device transconductance in the long-channel approximation, using the relationship:[17]

$$\mu = 2L_{DS}\, I_{DS} / [C_G\, W_{DS}\, (V_G - V_T)^2], \qquad (2)$$

where $L_{DS}$ and $W_{DS}$ are the D-S channel length and width, respectively, $C_G$ is the gate insulator specific capacitance ($C_G$ = 32.7 nF/cm$^2$ for 100-nm SiO$_2$ layers) and $V_T$ is the threshold voltage. Four different estimates were carried out for determining $\mu_{e/h}^{(j)}$ in both the "0" semiconducting and "1" semimetallic regimes, and from the branches of the transfer characteristics associated to minority (e) and majority (h) carriers, respectively.

**Acknowledgements**

This work was supported by a Strategic Partnership Grant (SPG-P) from the Natural Sciences and Engineering Research Council (NSERC) of Canada (GF & JBG: 506356-2017), and the Canada Foundation for Innovation (JBG: JELF-33977; GF: LOF-212442). GF acknowledges a Canada research Chair in Carbon-based nanomaterials and Nano-optoelectronics. The authors would like to thank Dr. Andranik Sarkissian at Plasmionique Inc. for providing the software for monitoring the device contacting system, and for technical support.

**Author information**

[*] Corresponding authors: *joe.gilroy@uwo.ca* and *gfanchin@uwo.ca*

*Supplementary Information*

**Transparent and flexible field-effect transistors and mem-transistors with electroactive layers of solution-processed organic polyradicals**

Deepa Singh[1,2], François Magnan[2,3], Joe B. Gilroy[2,3] *, and Giovanni Fanchini[1,2,3] *

[1]Department of Physics and Astronomy, The University of Western Ontario, 1151 Richmond St., London, Ontario N6A 3K7, Canada.

[2]Centre for Advanced Materials and Biomaterials Research (CAMBR), The University of Western Ontario, 1151 Richmond St., London, Ontario, N6A 5B7, Canada.

[3]Department of Chemistry, The University of Western Ontario, 1151 Richmond St., London, Ontario, N6A 5B7, Canada.

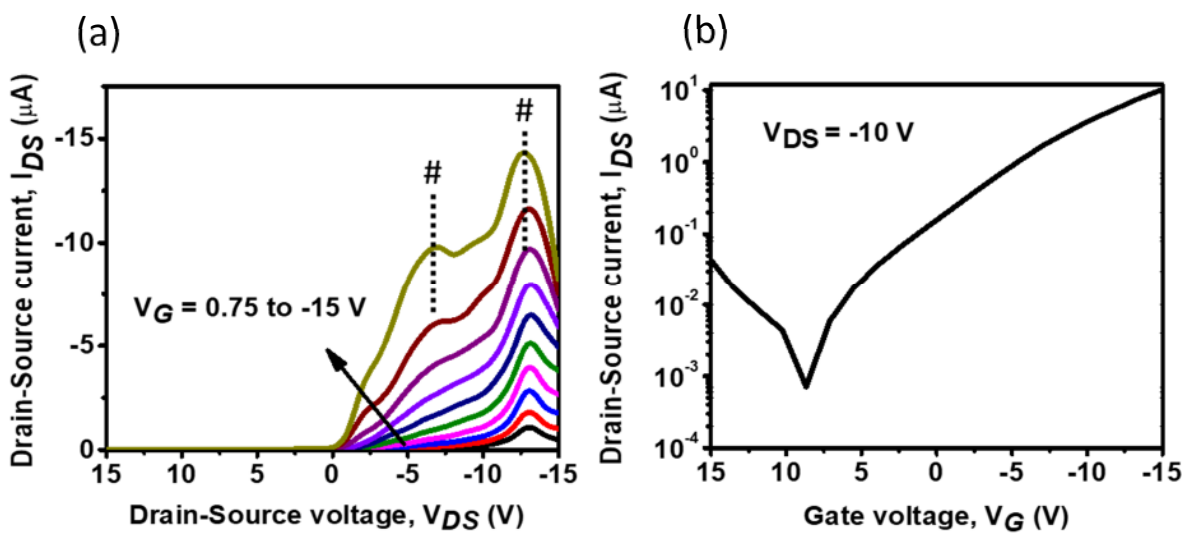

**Fig. S1 - a.** *Drain-source current ($I_{DS}$) vs. Drain-source voltage ($V_{DS}$) output characteristics and* **b.** *Gate transfer characteristics for the PR-FET spun from chlorobenzene (same as Fig. 3e) after thermal annealing at 120 °C in inert nitrogen atmosphere ($O_2$ level below 5 ppm and moisture below 2 ppm). It is observed that the current surges ("kinks", indicated with # in panel a) further increase upon annealing, a procedure designed to lead to the desorption of residual molecules from the processing solvent [S1] so they are assigned to inherent properties in P6OV and will be the subject of future investigations.*

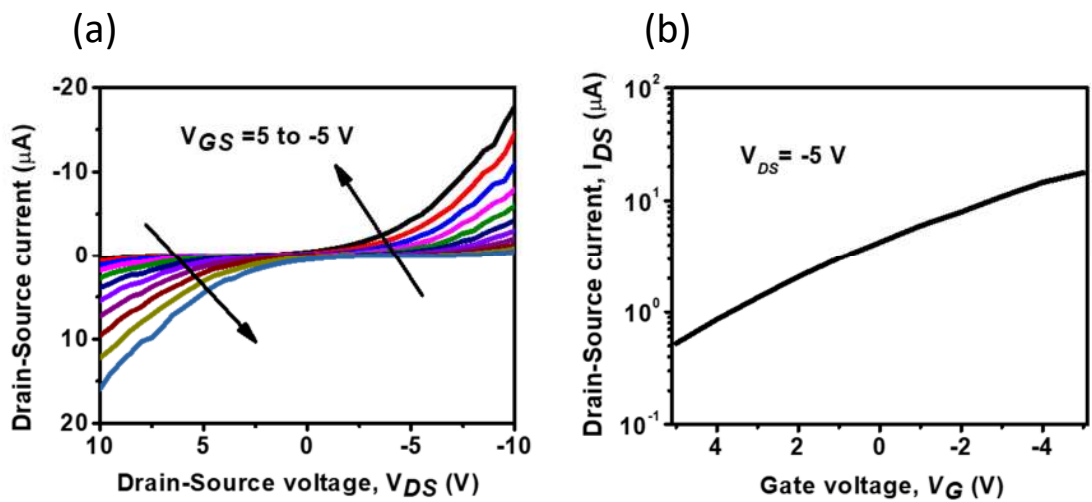

**Fig. S2 - a.** *Drain-source current ($I_{DS}$) vs. Drain-source voltage ($V_{DS}$) output characteristics and* **b.** *Gate transfer characteristics for the PR-FET with aluminum D-S contacts (as in Fig. 4b). It is observed that this PR-FET is in the ambipolar "1" charge transport regime, consistently with injection of comparable rates of both electrons and holes by Al in the P6OV active layer* [S2]